\documentclass[epj]{svjour}

\usepackage[dvips]{graphicx}

\usepackage{array}
\usepackage{amssymb}
\usepackage{amsmath}
\usepackage{hhline}
\usepackage{longtable}
\usepackage{dcolumn}
\usepackage{bm}
\usepackage{subfigure}
\usepackage{epsfig}
\usepackage{latexsym,amsmath}
\usepackage{amsbsy}

\makeatletter
\renewcommand\appendix{\par
  \setcounter{section}{0}%
  \setcounter{subsection}{0}%
  \setcounter{equation}{0}%
  \renewcommand\theequation{\Alph{section}.\arabic{equation}}
  \renewcommand\thesection{Appendix \@Alph\c@section:}}
\makeatother

\begin{document}
\title{Random Sierpinski network with scale-free small-world and modular structure}

\author{Zhongzhi Zhang\inst{1,2} \thanks{e-mail: zhangzz@fudan.edu.cn} \and Shuigeng Zhou\inst{1,2} \thanks{e-mail: sgzhou@fudan.edu.cn} \and Zhan Su\inst{1,2} \and Tao Zou\inst{1,2}  \and Jihong Guan\inst{3}}                     
\institute{Department of Computer Science and Engineering, Fudan
University, Shanghai 200433, China \and Shanghai Key Lab of
Intelligent Information Processing, Fudan University, Shanghai
200433, China \and Department of Computer Science and Technology,
Tongji University, 4800 Cao'an Road, Shanghai 201804, China}

\date{Received: date / Revised version: date}

\abstract{In this paper, we define a stochastic Sierpinski gasket,
on the basis of which we construct a network called random
Sierpinski network (RSN). We investigate analytically or numerically
the statistical characteristics of RSN. The obtained results reveal
that the properties of RSN is particularly rich, it is
simultaneously scale-free, small-world, uncorrelated, modular, and
maximal planar. All obtained analytical predictions are successfully
contrasted with extensive numerical simulations. Our network
representation method could be applied to study the complexity of
some real systems in biological and information fields.
} 

 \maketitle

\section{Introduction}

In the last few years, much attention has been paid to the study of
complex networks as an interdisciplinary subject~\cite{AlBa02}. It
is now established that network science is a powerful tool in the
analysis of real-life complex systems by providing intuitive and
useful representations for networked systems. Many real-world
natural and man-made systems have been examined from the perspective
of complex network theory. Commonly cited examples include the
Internet~\cite{FaFaFa99}, the World Wide Web~\cite{AlJeBa99},
metabolic networks~\cite{JeToAlOlBa00}, protein networks in the
cell~\cite{JeMaBaOl01}, co-author networks~\cite{Ne01a}, sexual
networks~\cite{LiEdAmStAb01}, to name but a few. The empirical
studies have uncovered the presence of several generic properties
shared by a lot of real systems: power-law degree
distribution~\cite{BaAl99}, small-world effect including small
average path length (APL) and high clustering
coefficient~\cite{WaSt98}, and community (modular) structure~\cite
{GiNe02}. These new discoveries have inspired researchers to develop
a variety of techniques and models in an effort to understand or
predict the behavior of real systems~\cite{AlBa02}. It is still of
current interest to reveal other different processes in real-life
systems that may lead to above general characteristics.

In our earlier paper, we have proposed a family of deterministic
networks based on the well-known Sierpinski
fractals~\cite{ZhZhZoChGu07}. These networks posses good topological
properties observed in some real systems. However, their
deterministic construction are not in line with the randomness of
many real-world systems. In this paper, we present a stochastic
Sierpinski gasket, in relation to which a novel network, named
random Sierpinski network (RSN), is constructed. The obtained
network is a maximal planar graph, it displays the general
topological features of real systems: heavy-tailed degree
distribution, small-world effect, and modular structure. We also
obtain the degree correlations of RSN. All theoretical predictions
are successfully confirmed by numerical simulations.

\section{Brief introduction to Sierpinski network}

\begin{figure}
\begin{center}
\includegraphics[width=0.45\textwidth]{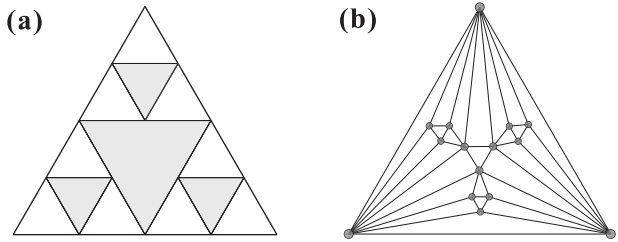}
\end{center}
\caption[kurzform]{\label{network} The first two stages of
construction of the Sierpinski gasket (a) and its corresponding
network (b).}
\end{figure}

In our previous work~\cite{ZhZhZoChGu07}, motivated by the classic
deterministic fractal, Sierpinski gasket (or Sierpinski
triangle)~\cite{Si15,Re94}, we introduce a new type of graph, called
Sierpinski network. The well-known Sierpinski gasket, shown in
figure~\ref{network}(a), is constructed as follows~\cite{Re94}: We
start with an equilateral triangle, and denote this initial
configuration by generation $t=0$. Then in the first generation
$t=1$, the three sides of the equilateral triangle are bisected and
the central triangle removed. This forms three copies of the
original triangle, and the procedure is repeated indefinitely for
all the new copies. In the limit of infinite $t$ generations, we
obtain the famous Sierpinski gasket, whose Hausdorff dimension is
$d_f=\ln3/\ln2$~\cite{Hu81}.

From this famous fractal we have defined the Sierpinski
network~\cite{ZhZhZoChGu07} as illustrated in
figure~\ref{network}(b). The translation from the fractal to network
generation is quite straightforward. Let each of the three sides of
a removed triangle correspond to a node (vertex) of the network and
make two nodes connected if the corresponding sides contact one
another. For uniformity, the three sides of the initial equilateral
triangle at step 0 also correspond to three different nodes. The
resultant Sierpinski network has a power-law degree distribution
$P(k)\sim k^{-\gamma}$ with $\gamma=1+\frac{\ln3}{\ln2}$, displays
small-world effect, and is disassortative.

\section{Random Sierpinski network and its iterative algorithm}
\begin{figure}
\begin{center}
\includegraphics[width=0.4\textwidth]{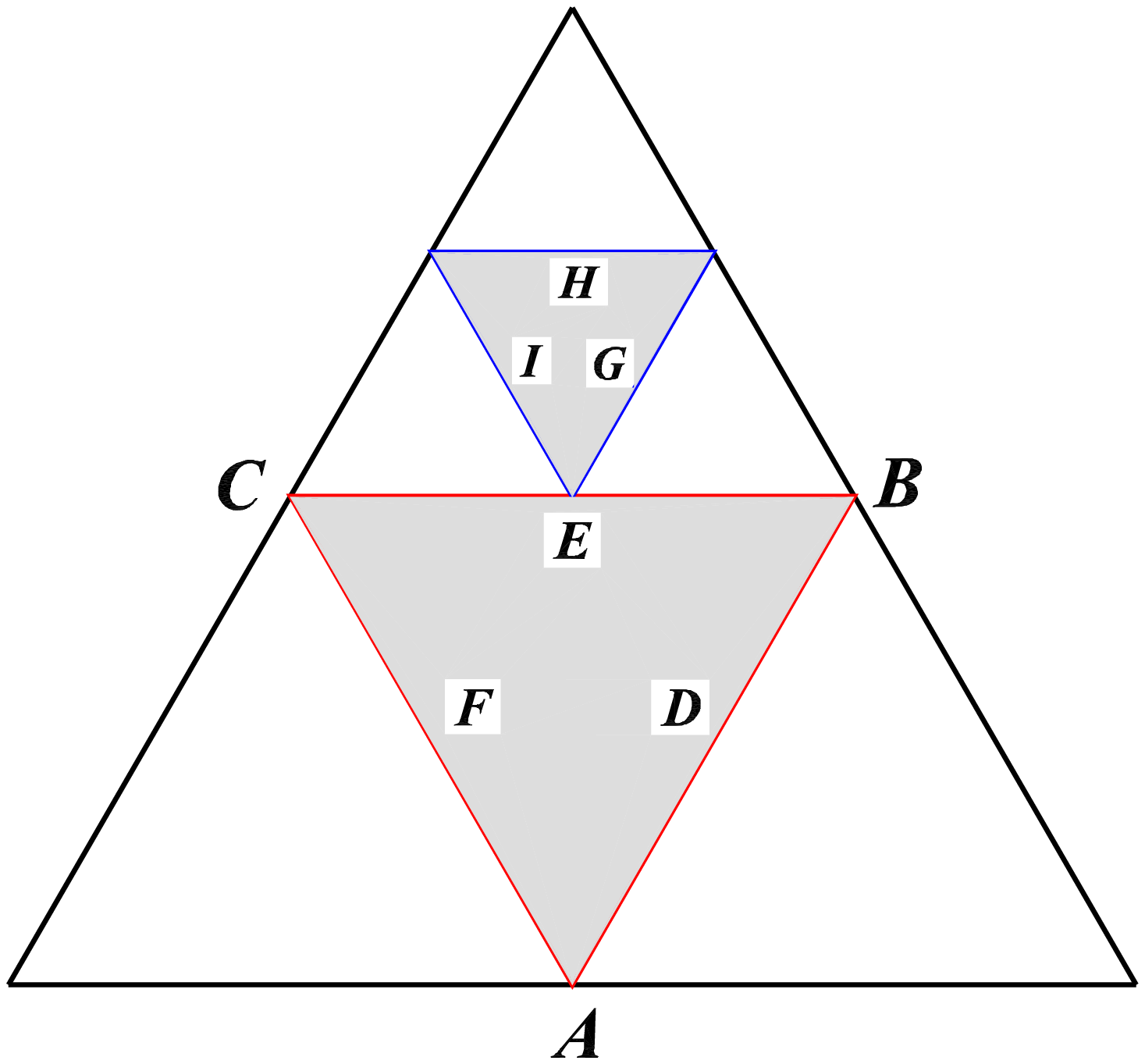}
\includegraphics[width=0.36\textwidth]{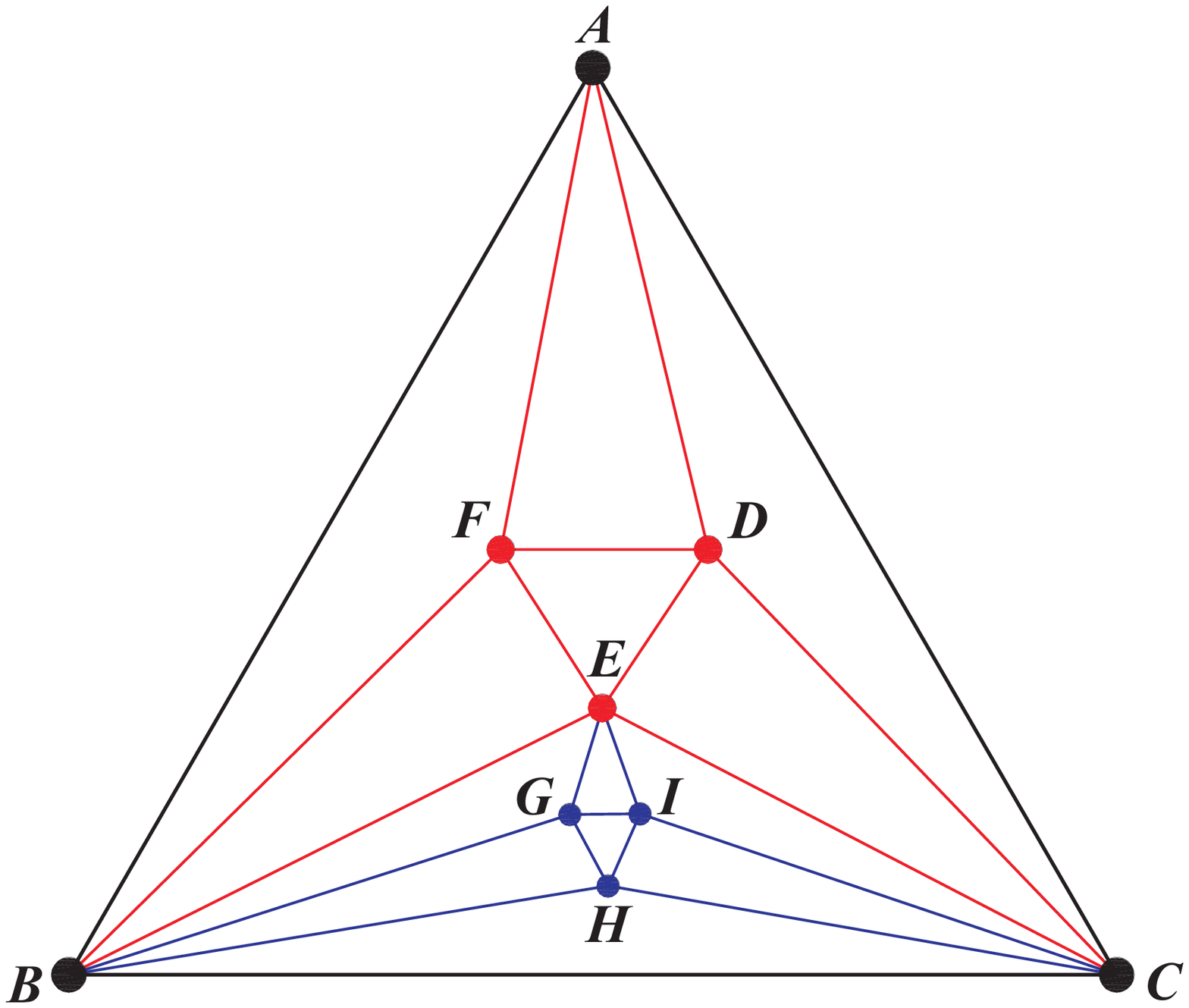}
\end{center}
\caption[kurzform]{\label{web} (Color online) The sketch maps for
the construction of random Sierpinski gasket (upper panel) and its
corresponding network (lower panel).}
\end{figure}

In this section, we first construct a random Sierpinski gasket from
the deterministic Sierpinski gasket (or Sierpinski
triangle)~\cite{Re94}. Then we will establish a random Sierpinski
network based on the proposed stochastic fractal. Analogous to the
Sierpinski triangle, the random Sierpinski gasket also starts with
an equilateral triangle. At step 1, we perform a bisection of the
sides and remove the downward pointing triangle forming three small
copies of the original triangle. Then in each of the subsequent
generations, an equilateral triangle is chosen randomly, for which
bisection and removal are performed to form three small copies of
it. The sketch map for the random fractal is shown in the upper
panel of figure~\ref{web}. From this fractal we can easily construct
the random Sierpinski network with sides of the removed triangles
mapped to nodes and contact to links between nodes. As in the
construction of the deterministic version~\cite{ZhZhZoChGu07}, the
three sides of the initial equilateral triangle at step 0 are also
mapped to three different nodes. Figure~\ref{web} (lower panel)
shows a network derived from the random Sierpinski gasket.

\begin{figure}
\begin{center}
\includegraphics[width=0.35\textwidth]{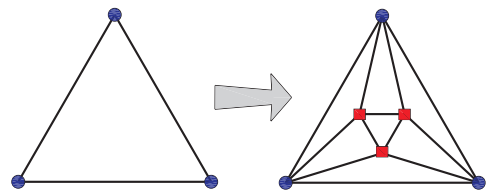}
\end{center}
\caption[kurzform]{\label{iterative}(Color online) Iterative
construction method for the network. }
\end{figure}

According to the construction of random Sierpinski network, we
introduce a general iterative algorithm generating the network. We
denote the random Sierpinski network after $t$ iterations by $W(t)$,
$t\geq 0$. Initially ($t=0$), $W(0)$ has three nodes forming a
triangle. At step $t=1$, we add three nodes into the original
triangle. These three new nodes are connected to one another shaping
a new triangle, and both ends of each edge of the new triangle are
linked to a node of the original triangle. Thus we obtain $W(1)$.
For $t\geq1$, $W(t)$ is obtained from $W(t-1)$. For the convenience
of description, we give the following definition: For each of the
existing triangles in $W(t-1)$, if there is no nodes in its interior
and among its three nodes there is only one youngest node (i.e., the
other two are strictly elder than it), we call it an \emph{active
triangle}. At step $t-1$, we select at random an existing active
triangle and replace it by the connected cluster on the right of
figure~\ref{iterative}, then $W(t)$ is produced.

Since at each time step the numbers of the nodes and edges increase
by 3 and 9, respectively, we can easily know that at step $t$, the
network consists of $N_t=3t+3$ nodes and $E_t=9t+3$ edges. Thus, the
relation $E_t=3N_{t}-6$ holds for all steps. In addition, according
to the connection rule, arbitrary two edges in the network never
cross each other. Therefore, the considered network is a maximal
planar graph~\cite{We01}, which is similar to its deterministic
version~\cite{ZhZhZoChGu07} and some previously studied
networks~\cite{AnHeAnSi05}.

\section{Structural characteristics}
We now study the statistical properties of RSN, in terms of degree
distribution, clustering coefficient, average path length,
degree-degree correlations, and modularity. The analytical
approaches are completely deferent from those applied to the
deterministic Sierpinski network~\cite{ZhZhZoChGu07}.

\subsection{degree distribution}
Initially ($t=0$), there is only one active triangle in the network.
In the subsequent iterations, at each time step, three active
triangles are created and one active triangle is deactivated
simultaneously, so the total number of active triangles increases by
2. Then at time $t$, there are $2t+1$ active triangles in RSN. Note
that, for an arbitrary given node, when it is born, it has a degree
of 4 and one active triangle containing itself; and in the following
steps, each of its two new neighbors separately generates a new
active triangle involving it, and one of its existing active
triangles is deactivated at the same time. So, for a node with
degree $k$, the number of active triangles containing it is
$\frac{k-2}{2}$. Let $N_k(t)$ denote the average number of nodes
with degree $k$ at time $t$. By the very construction of RSN, the
rate equation that accounts for the evolution of $N_k(t)$ with time
$t$ is~\cite{KaReLe00}
\begin{equation}\label{rate}
\frac{dN_k(t)}{dt}=\frac{\frac{k-4}{2} N_{k-2}(t)-\frac{k-2}{2}
N_k(t)}{2t+1}+3\,\delta_{k,4}.
\end{equation}
The first term on the right-hand side (rhs) of Eq.~(\ref{rate})
accounts for the process in which a node with $k-2$ links is
connected to two new nodes, leading to a gain in the number of nodes
with $k$ links. Since there are $N_{k-2}(t)$ nodes of degree $k-2$,
such processes occur at a rate proportional to
$\frac{k-4}{2}N_{k-2}(t)$, while the factor $2t+1$ converts this
rate into a normalized probability. A corresponding role is played
by the second (loss) term on the rhs of Eq.~(\ref{rate}). The last
term on the rhs of Eq.~(\ref{rate}) accounts for the continuous
introduction of three new nodes with degree four.

In the asymptotic limit $N_k(t)=(3t+3)P(k)$, where $P(k)$ is the
degree distribution. Substitute this relation into Eq.~(\ref{rate})
to lead to the following recursive equation for infinite $t$
\begin{equation}\label{recursive}
P(k)=\left\{\begin{array}{lcl}
\frac{k-4}{k+2}P(k-2) & \mbox{for} & k\geq 4+2\\
\frac{2}{3} & \mbox{for} & k=4\,,\\
\end{array}\right.
\end{equation}
giving
\begin{equation}\label{degree}
P(k)=\frac{32}{k(k+2)(k-2)}.
\end{equation}
In the limit of large $k$, $P(k)\sim k^{-3}$, which has the same
degree exponent as the BA model~\cite{BaAl99} and some hierarchical
lattice models~\cite{HiBe06}.
\begin{figure}
\begin{center}
\includegraphics[width=.3\linewidth, trim=100 10 140 20]{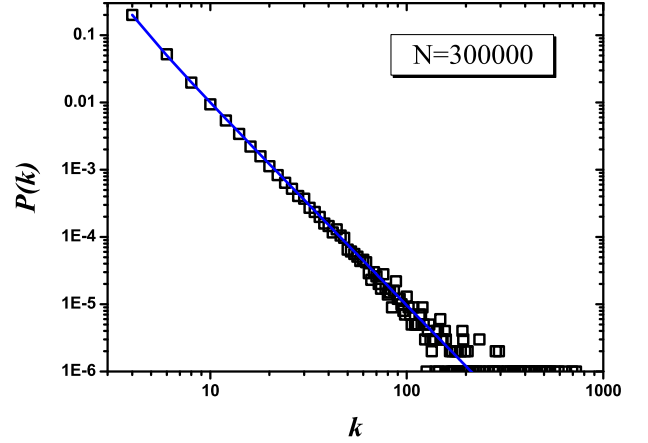}
\caption{(Color online) Log-log graph of the degree distribution for
a network with order $N=300000$. The squares denote the numerical
results and the solid line shows the theoretical predication given
by Eq.~(\ref{degree}).} \label{DegDistri}
\end{center}
\end{figure}
In order to confirm the analytical prediction, we performed
numerical simulations of the network plotted in
figure~\ref{DegDistri}, which shows that the simulation result is
well in agreement with the analytic one.

\subsection{clustering coefficient}

By definition, the clustering coefficient~\cite{WaSt98} $ C_i $ of
node $i$ is defined as the ratio between the number of edges $e_i$
that actually exist among the $k_i $ neighbors of node $i$ and its
maximum possible value, $ k_i( k_i -1)/2 $, i.e., $ C_i =2e_i/[k_i(
k_i -1)]$. Generally, in a network, for nodes with degree $k$, their
clustering coefficients, $C(k)$, are not always the same. But in our
network, all nodes with the same degree have identical clustering
coefficient. Moreover, for a single node with degree $k$, the
analytical expression for its clustering coefficient $C(k)$ can be
derived exactly.

According the connection rule (see figure~\ref{iterative}), when a
node $i$ enters the system, both $k_i$ and $e_i$ are 4. In the
following steps, if one of its active triangles is selected, both
$k_{i}$ and $e_{i}$ increase by 2 and 3, respectively. Thus, $e_{i}$
equals to $4+\frac{3}{2}\left(k_{i}-4\right)$. The relation holds
for all nodes at all steps. So one can see that there exists a
one-to-one correspondence between the degree of a node and its
clustering. For a node of degree $k$, we have
\begin{equation}\label{Ck}
C(k)= \frac{2\,e}{k(k-1)}=
\frac{2\left[4+\frac{3}{2}(k-4)\right]}{k(k-1)}=\frac{4}{k}-\frac{1}{k-1}.
\end{equation}
In the limit of large $k$, $C(k)$ exhibits a power-law behavior,
$C(k)\sim k^{-1}$, which has also been empirically observed in
several real networks~\cite{RaBa03}.

We continue to compute the average clustering coefficient $C$ of RSN
by means of the clustering spectrum $C(k)$:
\begin{equation}\label{ACC1}
  C = \sum_k P(k)C(k),
\end{equation}
which can be easily obtained with respect to the degree distribution
$P(k)$ expressed by Eq. (\ref{degree}). The result is
\begin{eqnarray}\label{ACC2}
 C &=&\sum_{k} \frac{32}{k(k+2)(k-2)}
 \left(\frac{4}{k-1}-\frac{1}{k-1}\right)\nonumber \\
&=&\frac{32\ln2}{3}-\frac{4\pi^2}{3}+\frac{19}{3}\approx 0.5674.
\end{eqnarray}
\begin{figure}
\begin{center}
\includegraphics[width=7.5cm]{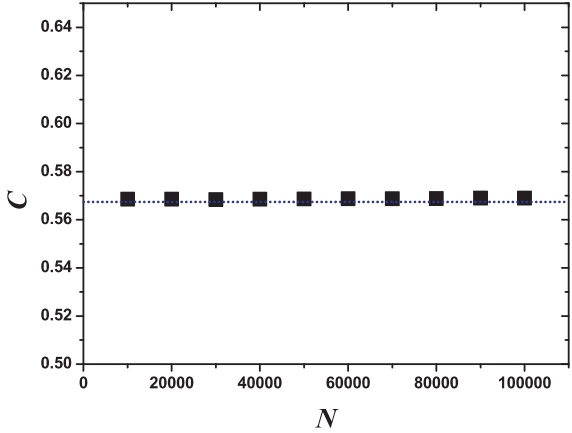}
\caption{(Color online) Average clustering coefficient $C$ of RSN vs
the network size $N$. The dotted line shows the analytic prediction
and the squares denote the simulation results. } \label{clu}
\end{center}
\end{figure}
Thus the average clustering coefficient $C$ of RAN is large and
independent of network size. We have performed extensive numerical
simulations of the RSN. In figure~\ref{clu}, we present the
simulation results about the average clustering coefficient of RSN,
which are in complete agreement with the analytical value.

\subsection{Average path length}

From above discussions, we find that the existing model shows both
the scale-free nature and the high clustering at the same time. In
fact, our model also exhibits small-word property. Next, we will
show that our network has at most a logarithmic average path length
(APL) with the number of nodes. Here APL means the minimum number of
edges connecting a pair of nodes, averaged over all couples of
nodes.

Using an mean-field approach similar to that presented in
Ref.~\cite{ZhYaWa05}, one can predict the APL of our network
analytically. By construction, at each time step, three nodes are
added into the network. In order to distinguish different nodes, we
construct a node sequence in the following way: when three new nodes
are created at a given time step, we label them as $M+1, M+2,\ldots,
M+3$, where $M$ is the total number of the pre-existing nodes.
Eventually, every node is labeled by a unique integer, and the total
number of nodes is $N_t=3t+3$ at time $t$.
\begin{figure}
\begin{center}
\includegraphics[width=7.5cm]{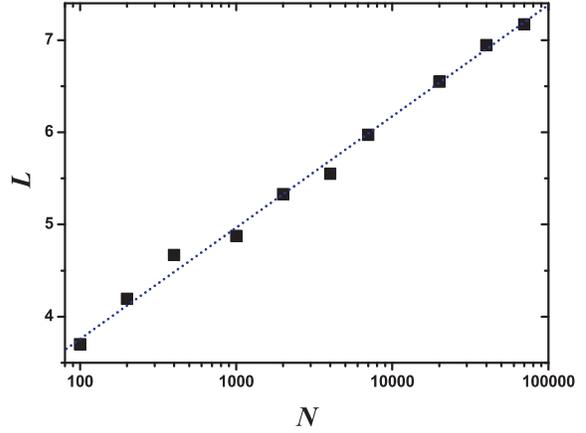}
\caption{(Color online) Semilogarithmic plot of the average path
length $L$ verse
 network size $N$. } \label{apl}
\end{center}
\end{figure}
We denote $L(N)$ as the APL of our network with size $N$. It follows
that $L(N)=\frac{2\,D(N)}{N(N-1)}$, where $D(N)=\sum_{1 \leq i<j
\leq N}d_{i,j}$ is the total distance, and where $d_{i,j}$ is the
smallest distance between node $i$ and $j$. Note that the distances
between existing node pairs are not affected by the addition of new
nodes. As in the analysis of~\cite{ZhYaWa05}, we can easily derive
that $D(N) \sim N^2\ln N$ in the infinite limit of $N$. Then, $L(N)
\sim \ln N$. Thus, there is a slow growth of the APL with the
network size $N$. This logarithmic scaling of $ L(N)$ with network
size $N$, together with the large clustering coefficient obtained in
the preceding subsection, shows that the considered graph has a
small-world effect. In figure~\ref{apl}, we report average path
length $L(N)$ versus network size $N$. One can obviously see that
$L(N)$ increases logarithmically with $N$.

\subsection{degree correlations}
First, we study the time evolution for the connectivity of an
arbitrary node. Notice that the growing precess of RSN actually
contains the preferential attachment mechanism, which arises in it
not because of some special rule including a function of degree as
in Ref.~\cite{BaAl99} but naturally. Indeed, the probability that
new nodes created at time $t$ will be connected to an existing node
$i$ is clearly proportional to the number of active triangles
containing $i$, i.e. to its $\frac{k_i(t)-2}{2}$. Thus a node $i$ is
selected with the usual preferential attachment probability
$\Pi_{i}[k_i(t)]=[k_i(t)-2]/[2(2t+1)] \sim [k_i(t)-2]/4t$ (for large
$t$). Consequently, $k_i$ satisfies the dynamical equation
\cite{AlBa02}:
\begin{equation}
\frac{\partial k_i(t)}{\partial t} =2\cdot\frac{k_i(t)-2}{4t}.
\end{equation}
Considering the initial condition $k_i(t_i)=4$, we have
\begin{equation}\label{connect}
k_i(t)=2\left(\frac {t}{i}\right)^{1/2}+2.
\end{equation}

Having obtained the degrees for all nodes, we now study the degree
correlations. Generally, degree correlations in a network can be
conveniently measured by means of the quantity, called \emph{average
nearest-neighbor degree} (ANND), which is a function of node degree,
and is more convenient and practical in characterizing degree-degree
correlations.  The ANND is defined by~\cite{PaVaVe01}
\begin{equation}
  k_{nn}(k) = \sum_{k'} k' P(k'|k),
  \label{knn1}
\end{equation}
where $P(k'|k)$ is the condition probability that a link belonging
to a node with connectivity $k$ points to a node with connectivity
$k'$. If there are no two degree correlations, $k_{nn}(k)$ is
independent of $k$. When $k_{\rm nn}(k)$ increases (or decreases)
with $k$, the network is said to be assortative (or
disassortative)~\cite{Newman02}.

Correlations can also be described by a Pearson correlation
coefficient $r$, which is defined as~\cite{Newman02}:
\begin{equation}\label{assort}
 r=
 \frac{
\frac{1}{M}\sum_{m}j_mk_m-\left [\frac{1}{M}
\sum_{m}{\frac{1}{2}(j_{m}+k_{m})}\right ]^2} {\frac{1}{M}
\sum_{m}{\frac{1}{2}(j_m^2+k_m^2)} - \left [ \frac{1}{M}
\sum_{m}{\frac{1}{2}(j_{m}+k_{m})}\right ]^2}\,,
\end{equation}
where $j_{m}$, $k_{m}$ are the degrees of the vertices at the ends
of the $m$th edge, with $m = 1,2,\cdots, M$, where $M$ denotes the
number of edges in the network. The coefficient is in the range
$-1\leq r \leq 1$. If the network is uncorrelated, the correlation
coefficient equals zero. Disassortative networks have $r<0$, while
assortative graphs have a value of $r>0$.

We can analytically calculate the function value of $k_{nn}(k)$ for
the RSN. Let $R_i(t)$ denote the sum of the degrees of the neighbors
of node $i$, evaluated at time $t$. It is represented as
\begin{equation}
  R_i(t) =  \sum_{j \in \Omega(i)} k_j(t),
  \label{Ri}
\end{equation}
where $\Omega(i)$ corresponds to the set of neighbors of node $i$.
The ANND of node $i$ at time $t$, $k_{nn}(i,t)$, is then given by
$k_{nn}(i,t) = R_i(t) / k_i(t)$. During the growth of the RSN,
$R_i(t)$ can only increase by the addition of new nodes connected
either directly to $i$, or to one of the neighbors of $i$. In the
first case $R_i(t)$ increases by 8 (the sum of degree for two
newly-created nodes), while in the second case it increases by 2.
Therefore, in the continuous $k$ approximation, we can write down
the following rate equation~\cite{BoPa05}:
\begin{eqnarray}\label{dRi}
  \frac{dR_i(t)}{dt}& = &8\, \Pi_{i} [k_i(t)] + 2\,\sum_{j \in \Omega(i)}  \Pi_{j}
  [k_j(t)]\nonumber \\
  & =& \frac{2[k_i(t)-2]}{t} + \frac{R_i(t)-2k_i(t)}{2t} \nonumber \\
  & =& \frac{R_i(t)}{2t}+\frac{k_i(t)}{t}-\frac{4}{t}.
\end{eqnarray}
The general solution of Eq. (\ref{dRi}) is
\begin{equation}
  R_i(t)= 4 + \Phi_{0}(i) t^\frac{1}{2} +
  2 \left( \frac{t}{i} \right)^{1/2} \ln t,
  \label{eq:12}
\end{equation}
where $\Phi_{0}(i)$ is determined by the boundary condition
$R_i(i)$. To obtain the boundary condition $R_i(i)$, we observe that
at time $i$, the new node $i$ is connected to an existing node $j$
of degree $k_j(i)$ with probability $\Pi_{j}[k_j(i)]$, and that the
degree of this node increase by 2 in the process. Thus,
\begin{equation}
  R_i (i) = \sum_{j=1}^{i}\Pi_{j} [k_j(i)] [k_j(i)+2] + 8,
  \label{eq:2}
\end{equation}
where the last term 8 denotes the sum of the other two new nodes
created at the same time as node $i$. Inserting $\Pi_{j}
[k_j(i)]=\frac{k_j(i)-2}{4i}$ and $k_j(i)=2\left(\frac
{i}{j}\right)^{1/2}+2$ into $R_i(i)$ leads to
\begin{equation}
  R_i (i) =  8 + \ln i + \sum_{j=1}^{i} 2(ij)^{-\frac{1}{2}} \leq  8 + 3\ln i.
  \label{eq:11}
\end{equation}
So, in the large $i$ limit, $R_i (i)$ is dominated by the second
term, yielding
\begin{equation}
  R_i (i) \lessapprox 3\ln i.
  \label{qq}
\end{equation}
From here, we have
\begin{equation}
  R_i(t) \simeq 2 \left(\frac {t}{i}\right)^{1/2} \ln t,
\end{equation}
and finally
\begin{equation}
  k_{nn}(k,t) \simeq \ln t.
\label{eq:knnBA}
\end{equation}
So, two node correlations do not depend on the degree. The ANND
grows with the network size $N\approx 3t$ as $\ln N$, in the same
way as in the Barab\'asi-Albert (BA) model~\cite{VapaVe02} and the
two-dimensional random Apollonian network~\cite{ZhZh07}.
\begin{figure}
\begin{center}
\includegraphics[width=.3\linewidth, trim=120 10 120 20]{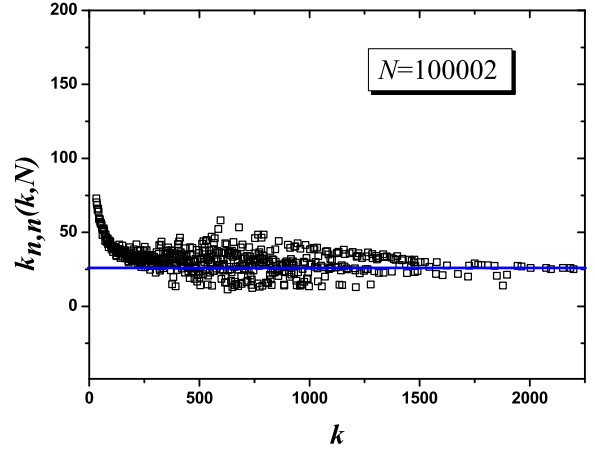}
\caption{(Color online) Plot of average nearest-neighbor degree of
the nodes with degree $k$. The squares denote the simulation
results, while the solid line is the theoretical result provided by
Eq.~(\ref{eq:knnBA}).} \label{ANND}
\end{center}
\end{figure}
In order to confirm the validity of the obtained analytical
prediction of ANND, we performed extensive numerical simulations of
the RSN (see figure~\ref{ANND}) with order $N=100002$. To reduce the
effect of fluctuation on simulation results, the simulation results
are average over fifty network realizations. From figure~\ref{ANND}
we observe that for large $k$ the ANND of numerical and analytical
results are in agreement with each other, while the simulated
results of ANND of small $k$ have a very weak dependence on $k$,
which is similar to the phenomena observed in the BA
model~\cite{VapaVe02}. This $k$ dependence, for small degree, cannot
be detected by rate equation approach, since it has been formulated
in the continuous degree $k$ approximation.

\begin{figure}
\begin{center}
  \includegraphics[width=7.5cm]{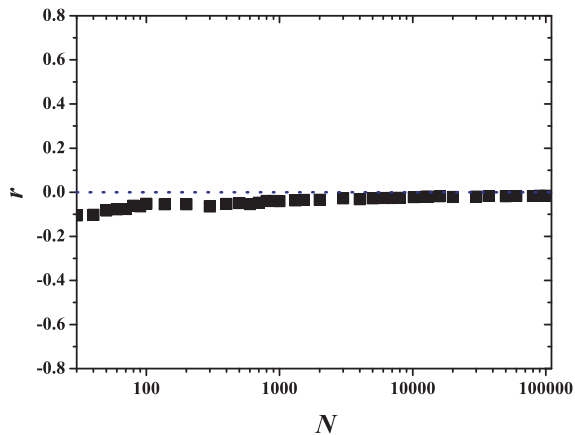}
  \caption{(Color online) Semilogarithmic graph of Pearson correlation coefficient $r$ as a function of network size $N$.} \label{pearson}
\end{center}
\end{figure}

To further confirm that RSN is uncorrelated, we compute the Pearson
correlation coefficient $r$ according to Eq.~(\ref{assort}). The
numerical results are reported in figure~\ref{pearson}. From this
figure, we see that for networks with small size, $r$ is negative
and only a little smaller than zero; when the size of the network
increases, $r$ goes to zero and is independent of size. The
phenomenon of the convergence of $r$ to zero again indicates that
RSN shows absence of degree correlations.

The fact that there is no degree correlation in RSN is compared with
that of its deterministic variant, which shows a negative degree
correlation~\cite{ZhZhZoChGu07}. We guess that the reason behind
this difference between RSN and its deterministic counterpart may
stem from a biased choice of ``active triangles". In the evolution
process of RSN, only one active triangle is updated at each
iteration, while for the deterministic version, all active triangles
are updated in one iteration. The difference between asynchronous
and synchronous updating leads to different power-law exponents of
degree distribution for the two networks, see~\cite{CoRobA05} for
detailed explanation. We think that the different assortative nature
between the two graphs might be also related to a biased choice of
active triangles at each iteration. Of course, the genuine reason
for this difference requires further study.


\subsection{modularity}

Many social and biological networks are fundamentally
modular~\cite{GiNe02,RaSoMoOlBa02,RaCaCeLoPa04}. These networks are
formed by communities (modules) of nodes that are highly
interconnected with each other, but have only a few or no links to
nodes outside of the community to which they belong to. The strength
of community structure is quantified through the
modularity~\cite{NeGi04}
\begin{equation}
Q =
\sum_{s=1}^{N_q}\left[\frac{l_s}{E}-\left(\frac{d_s}{2E}\right)^2\right],
  \label{Ri}
\end{equation}
where the sum runs over all communities, $N_q $ is the number of
communities (modules), $E$ is the link number in the network, $l_s$
is the total number of links in the $s$th community, and $d_s$ is
the sum of the connectivities (degrees) of the nodes in module $s$.
The modularity is high if the number of within- community links is
much larger than expected from chance alone.

We now look at the community structure of the network using the
algorithm originally proposed by Girvan and Newman (GN) to find the
partition with the largest modularity~\cite{GiNe02}. The GN
algorithm works by beginning with the complete network and at each
step removing the edge with largest betweenness, where this quantity
is recalculated after the removal of every edge. If there is more
than one edge with the same largest betweenness, we remove them all
at the same step. After all edges are removed, the network breaks up
into $N_t$ communities (non-connected nodes).
\begin{figure}
\begin{center}
\includegraphics[width=.3\linewidth, trim=110 10 110 15]{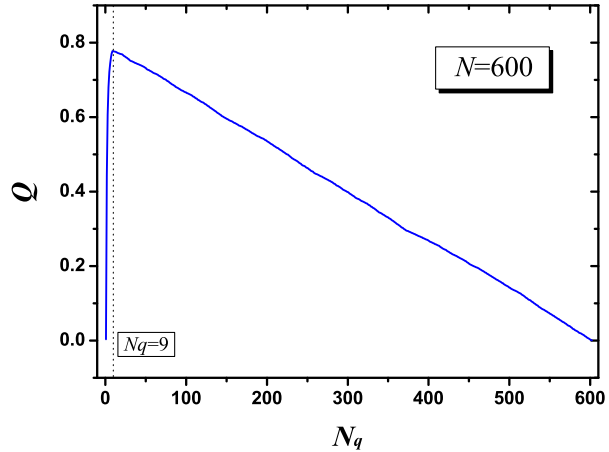}
\caption{(Color online) Modularity $Q$ as a function of number of
communities $N_q$. The vertical dashed line indicates the best split
with largest $Q$.} \label{modular}
\end{center}
\end{figure}
It is of interest to examine how the network is progressively broken
into separate communities as one removes more edges. In
figure~\ref{modular}, we present how $Q$ varies as the complete
network with size $N=600$ is broken up into communities. Obviously,
$Q$ has a broad large value as a function of the number of
communities, showing pronounced modular structure. $Q$ is greater
than 0.5 when there are between 3 and 230 communities, and the best
division has $N_q=9$ with $Q=0.7774$. This largest value of $Q$ is
in contrast to the highest values found for some real-life networks
such as collaboration network of scientists. Interestingly, the
combination of modular and uncorrelated properties has never been
reported in previous models.

\section{Conclusion}

In this paper, we have introduced a stochastic Sierpinski gasket and
related it to a random maximal network, called random Sierpinski
network (RSN). We have also proposed a iterative algorithm
generating RSN, based on which we have determined some relevant
topological characteristics of the network. We have presented that
the network is simultaneously scale-free, small-world, uncorrelated,
and modular. Thus, the RSN successfully reproduces some remarkable
properties of many natural and man-made systems. Our study provides
a paradigm of representation for the complexity of many real-life
systems, making it possible to study the complexity of these systems
within the framework of network theory.

\section*{Acknowledgment}

We thank Yichao Zhang for preparing this manuscript. This research
was supported by the National Basic Research Program of China under
grant No. 2007CB310806, the National Natural Science Foundation of
China under Grant Nos. 60496327, 60573183, 60773123, and 60704044,
the Shanghai Natural Science Foundation under Grant No. 06ZR14013,
the Postdoctoral Science Foundation of China under Grant No.
20060400162, Shanghai Leading Academic Discipline Project No. B114,
the Program for New Century Excellent Talents in University of China
(NCET-06-0376), and the Huawei Foundation of Science and Technology
(YJCB2007031IN).


\end{document}